\documentstyle[12pt,twoside,fleqn,espcrc1,epsfig]{article}

\newcommand{\beq}{\begin{equation}}
\newcommand{\eeq}{\end{equation}}
\newcommand{\bea}{\begin{eqnarray}}
\newcommand{\eea}{\end{eqnarray}}
\newcommand{\bce}{\begin{center}}
\newcommand{\ece}{\end{center}}
\newcommand{\etal}{{\it et al.}}

\newcommand{\ie}{{\it i.e.}}

\newcommand{\AmS}{{\protect\the\textfont2
  A\kern-.1667em\lower.5ex\hbox{M}\kern-.125emS}}

\title{Regeneration of Anti-Protons in Ultrarelativistic Heavy-Ion Collisions
\thanks{work supported by US-DOE grant DE-FG02-88ER40388.}} 

\author{Ralf Rapp and Edward V. Shuryak
\\ 
\vspace{0.3cm}
Department of Physics and Astronomy, SUNY at Stony Brook, NY 11794-3800, USA}

\begin{document}

% typeset front matter
\maketitle

\begin{abstract}
The production and annihilation of antiprotons in the hadronic phase of 
heavy-ion collisions is evaluated within a thermal equilibrium approach. 
It is shown that the inverse reaction of $p\bar p$ annihilation 
(\ie, multi-pion annihilation $N_\pi \pi\to p\bar p$ with $N_\pi\simeq$~5-7)
in connection with oversaturation of pion phase space (\ie, finite 
pion chemical potentials) plays an important 
role in understanding the observed antiproton yields at SpS energies
within the standard picture of subsequent chemical and thermal freezeout. 
Implications for RHIC energies are also addressed. 
\end{abstract}

%%%%%%%%%%%%%%%%%%%%%%%%%%%%%%%%%%%%%%%%%%%%%%%%%%%%%%%%%%%%%%%%%%%%%%
\section{Introduction}
%%%%%%%%%%%%%%%%%%%%%%%%%%%%%%%%%%%%%%%%%%%%%%%%%%%%%%%%%%%%%%%%%%%%%%
An important question in heavy-ion collisions at high energies concerns 
whether reinteractions between produced particles are frequent enough to 
establish and maintain thermal (and/or chemical) equilibrium, which 
constitutes an inevitable prerequisite to investigate the phase diagram 
of QCD. One way of addressing this issue is by testing predictions of 
equilibrium approaches with a large class of observables, aiming at a 
consistent description within one common scenario.   
In the context of the SpS heavy-ion experiments, this program has been 
carried out with remarkable success, as reflected by a simultaneous
description of hadron abundances~\cite{pbm99}, $p_t$-spectra encoding 
various patterns of hydrodynamic (collective) flow effects~\cite{HS98}, 
two-particle correlations~\cite{WH99}, etc..  
The deduced picture is that of a subsequent {\it chemical} and 
{\it thermal} freezeout, being characterized by the respective points 
$(\mu_N^{chem},T_{chem})\simeq (270,170)$~MeV~\cite{pbm99} and 
$(\mu_N^{therm},T_{therm})\simeq (410,120)$~MeV in terms of  
temperature and nucleon chemical potential coordinates of the phase diagram.
The dynamical justification resides in a hierarchy of hadronic interaction
strengths:
on the one hand, typical {\it elastic} cross sections are large
($\sim$~100~mb), supporting thermal equilibrium between $T_{chem}$ and 
$T_{therm}$ for about 5-10~fm/c; on the other hand, inelastic (number-changing) 
reactions have cross sections of 1-2 orders of magnitude smaller so 
that the net abundances of stable particles (nucleons, pions, kaons) are 
not significantly altered during the hadronic phase.   
Furthermore, the nontrivial features observed in electromagnetic spectra,
such as the enhancement found in both low- and intermediate-mass dilepton
as well as direct photon spectra, can also be accounted for within the same 
thermodynamic framework (cf.~ref.~\cite{RW00} for a recent review).   

In the following we investigate an observable -- antiproton yields --
that so far has been difficult to accommodate in this viewpoint of the 
collision dynamics, cf.~also ref.~\cite{RS00} and its application to 
antihyperons in ref.~\cite{GL00}.  
%thus raising doubts about its viability. 
%putting into doubt its correctness.  

%%%%%%%%%%%%%%%%%%%%%%%%%%%%%%%%%%%%%%%%%%%%%%%%%%%%%%%%%%%%%%%%%%%%%%
\section{Antiproton Production at CERN-SpS Energies}
\begin{figure}[!h]
\begin{minipage}[t]{7.5cm}
\vspace{-0.7cm}
\epsfig{file=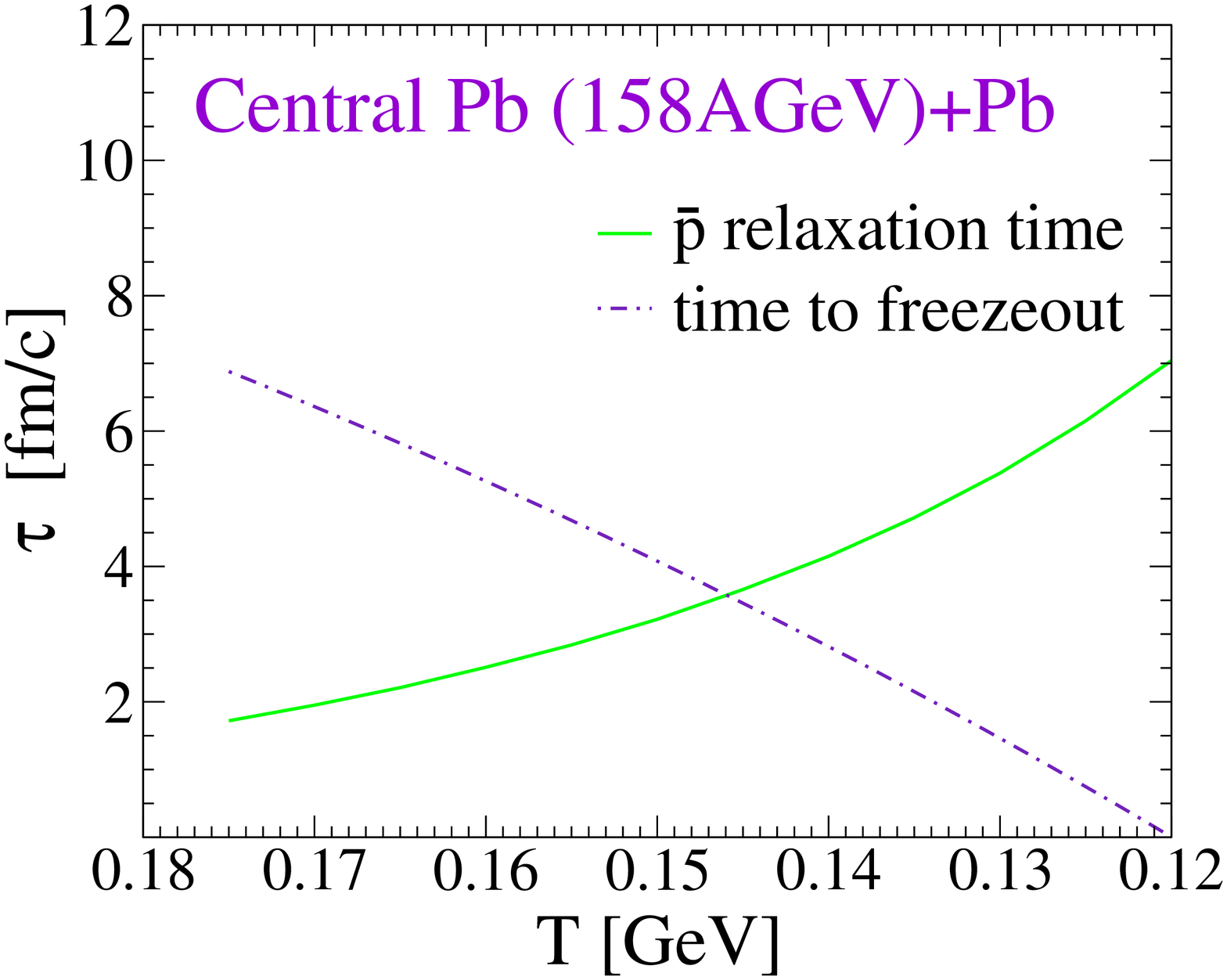,width=6.3cm,angle=-90}
\vspace{-1.2cm}
\caption{Chemical relaxation time for antiprotons (solid line), 
eq.~(\protect\ref{tauchem}), and remaining fireball lifetime until thermal
freezeout (dashed-dotted line) as obtained in the thermal model of 
ref.~\protect\cite{RW99}.}
\vspace{-0.5cm}
\label{fig_tauSpS}
\end{minipage}\hfill
\begin{minipage}[t]{7.5cm}
\vspace{-0.7cm}
\epsfig{file=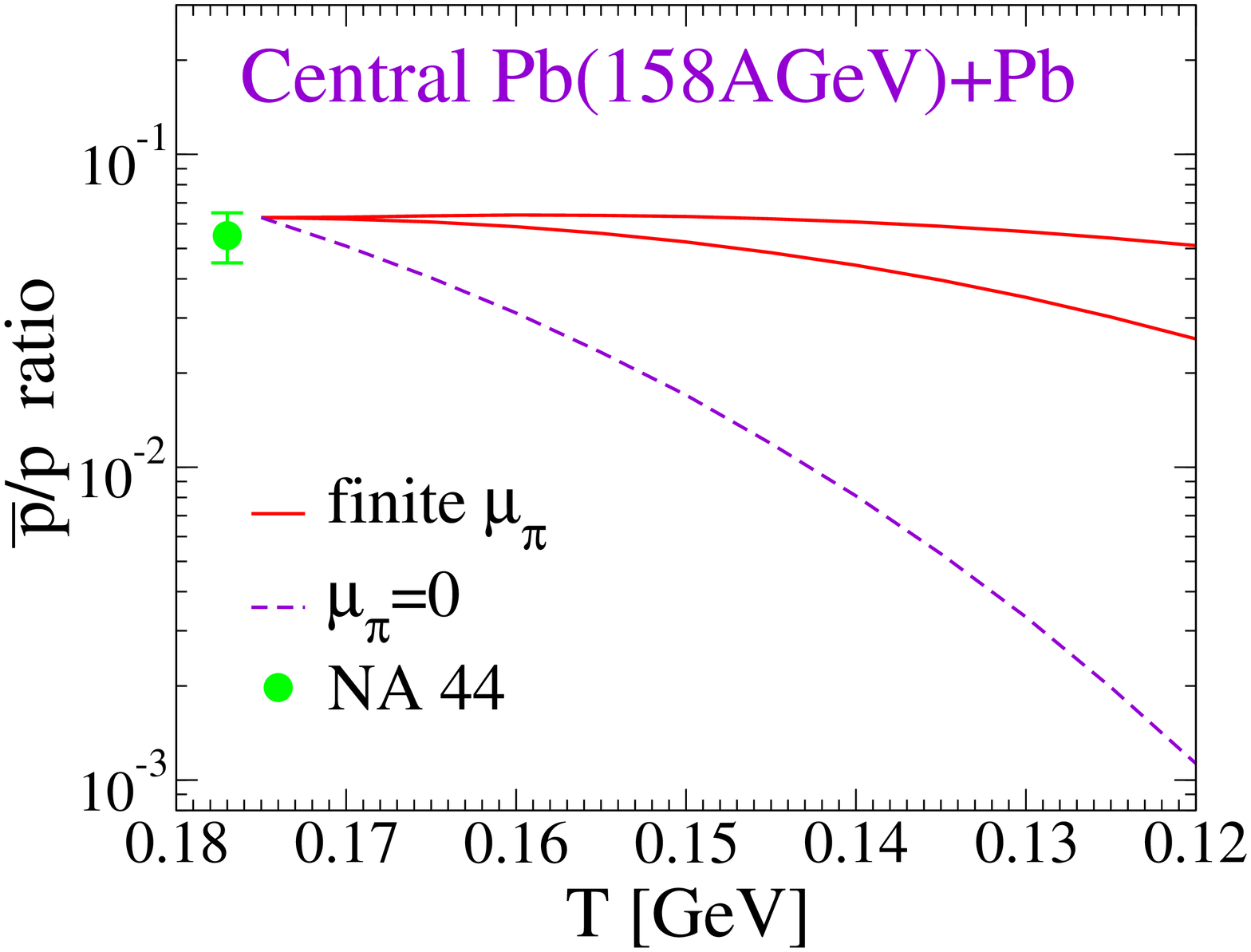,width=6.3cm,angle=-90}
\vspace{-1.2cm}
\caption{Temperature evolution of the $\bar p/p$ ratio with zero (dashed line)
and finite (solid lines) pion chemical potentials. The two solid lines reflect 
uncertainties corresponding to $\mu_\pi^{therm}$=65-80~MeV.} 
\vspace{-0.5cm}
\label{fig_ratioSpS}
\end{minipage}
\end{figure}

%%%%%%%%%%%%%%%%%%%%%%%%%%%%%%%%%%%%%%%%%%%%%%%%%%%%%%%%%%%%%%%%%%%%%% 
A copious production of antiprotons has been among early suggestions
for signals of Quark-Gluon Plasma formation in heavy-ion 
reactions~\cite{H86K88} (based on the much reduced threshold for 
anti-particle production in a partonic as compared to a hadronic 
environment). At the SpS the antiproton-to-proton ratio measured in 
central Pb(158~AGeV)-Pb collisions~\cite{na4x} is in agreement with the 
value predicted by the corresponding hadrochemical freezeout, 
$\bar p/p=\exp[-2\mu_N^{chem}/T_{chem}]\simeq$~5\%. The large annihilation 
cross section of antiprotons ($\sigma_{p\bar p}^{ann}\simeq$~50~mb at the relevant 
thermal energies of $\sqrt{s}\simeq$~2.3~GeV) renders this a rather  
puzzling fact, since the pertinent chemical relaxation time, 
\beq
\tau_{\bar p}^{chem}=\frac{1}{\sigma_{p\bar p}^{ann}(s_{th}) \ \varrho_B \ v_{th}} \ ,
\label{tauchem}
\eeq
stays below the fireball lifetime until rather late in the 
hadronic evolution, cf.~fig.~\ref{fig_tauSpS}. 
Indeed, standard hadronic transport 
calculations substantially underestimate the antiproton production. 
This led to speculations on various in-medium effects such as 
a shielding of the annihilation~\cite{ARC} or enhanced production via an 
increased string tension in the prehadronic stages~\cite{URQMD}. 
However, present transport approaches are not able to consistently treat 
reactions with more than 2 particles in the incoming channel.
Thus, although $p\bar p$ annihilation is included, the inverse reaction of 
multi-pion annihilation is not, entailing a violation of detailed balance. 
This observation alone, though, does not yet 
resolve the puzzle: a naive estimate assuming an equilibrium $\bar p$
abundance at thermal freezeout results in 
$\bar p/p = \exp[-2\mu_N^{therm}/T_{therm}]=0.1$~\%, a factor $\sim$50
below the measured value. What is missing here is that, after chemical
freezeout -- due to effective pion-number conservation -- the thermal pion 
densities exceed their chemical equilibrium values.
In statistical mechanics language this can be described by the build-up
of finite pion chemical potentials. The thermal rate equation for
$p\bar p \leftrightarrow N_\pi\pi$ then takes the form
\bea
{\cal R}_{th} = \int
d^3\tilde k_p~d^3\tilde k_{\bar p}~d^3\tilde k_{1} \cdots
d^3\tilde k_{N_\pi}  \delta^{(4)}(K_{tot}) ~|{\cal M}_{N_\pi}|^2
\{ z_p z_{\bar p} \ {\rm e}^{-(E_p+E_{\bar p})/T} -
z_\pi^{N_\pi} {\rm e}^{-\sum\limits_{i=1}^{N_\pi} \omega_{i}/T} \} \ ,
\label{rate}
\eea
where ${\cal M}_{N_\pi}$ is the invariant scattering matrix element. The
4-momentum conserving $\delta$-function forces the sum of proton
and antiproton energies $E_{p(\bar p)}$ to equal the sum of pion energies
$\omega_i$. Insisting on chemical equilibrium for the 
$p\bar p \leftrightarrow N_\pi \pi$ reaction thus provides an equation for the 
antiproton fugacity, $z_{\bar p}= z_\pi^{N_\pi} z_p$, implying large enhancement 
factors in the presence of a finite $\mu_\pi$ (note that changes in $z_{\bar p}$ 
have negligible feedback on $z_\pi$ or $z_p$ under SpS conditions).
A more detailed calculation~\cite{RS00} including measured pion-multiplicity 
distributions and $cms$-energy dependencies in $p\bar p$ annihilation~\cite{Dov92}
yields an average fugacity factor   
\beq
\langle z_\pi^{N_\pi}\rangle=
\sum\limits_{N_\pi=2}^{N_{\pi}^{max}} w_{N_\pi} \ \exp[N_\pi \mu_\pi/T] \ 
\eeq
with $w_{N_\pi}$ the probability weight of the $N_\pi$-pion channel.
The temperature dependence of the ratio 
$\bar p/p=\langle z_\pi^{N_\pi}(T)\rangle \exp[-2\mu_N^{chem}(T)/T_{chem}]$ 
in relative chemical equilibrium is displayed in fig.~\ref{fig_ratioSpS} based on the 
thermal fireball model of ref.~\cite{RW99}. As a result of large enhancement factors, 
the antiproton abundance at $T_{chem}$ can essentially be supported towards thermal 
freezeout (where $\langle z_\pi^{N_\pi}\rangle\simeq$~25-50), 
thus demonstrating the importance of multi-pion back-reactions in the presence 
of pion oversaturation.  
 
%%%%%%%%%%%%%%%%%%%%%%%%%%%%%%%%%%%%%%%%%%%%%%%%%%%%%%%%%%%%%%%%%%%%%%
\section{Antiproton Production at RHIC Energies}
\begin{figure}[!h]
\begin{minipage}[t]{7.5cm}
\vspace{-0.7cm}
\epsfig{file=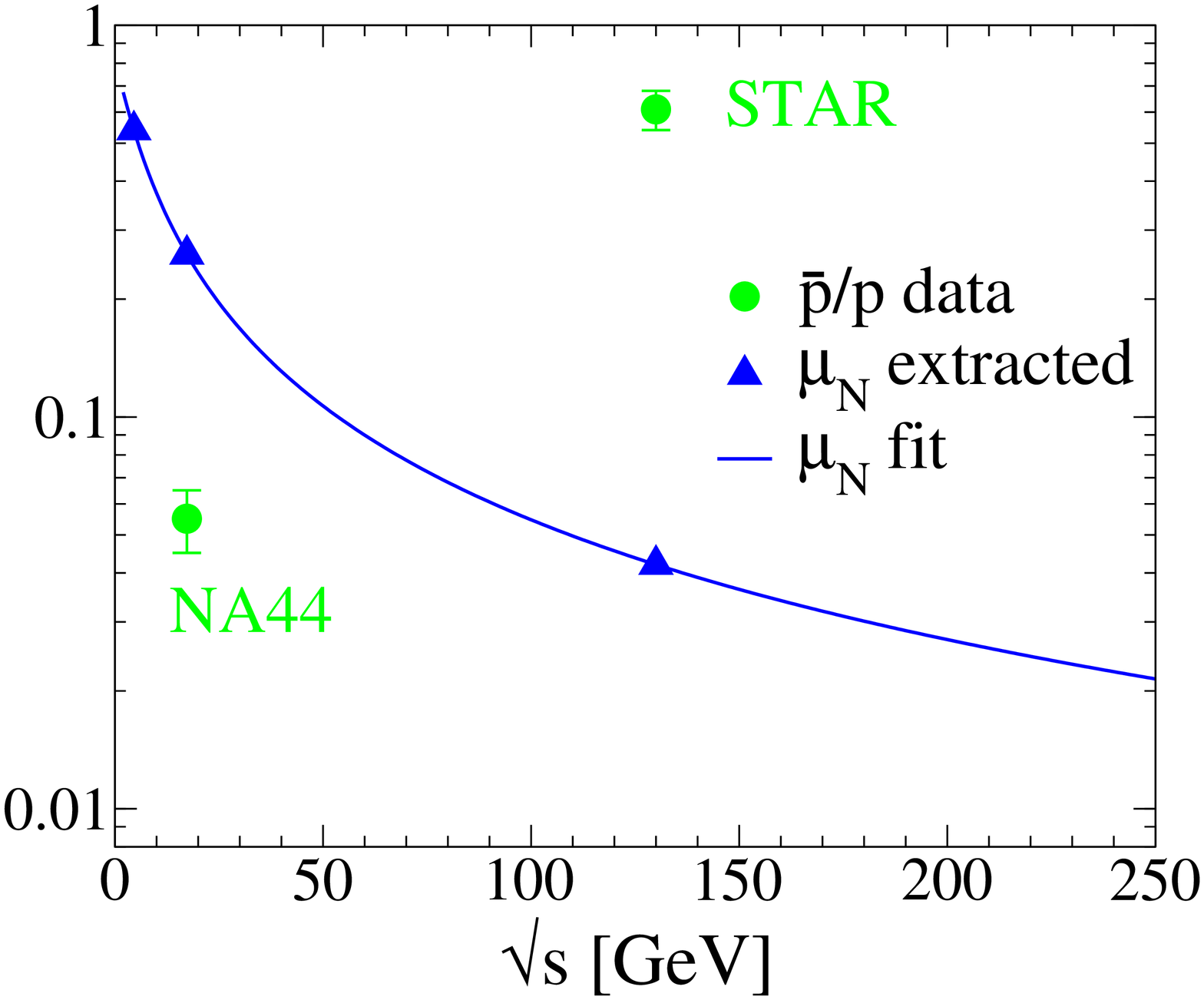,width=6.3cm,angle=-90}
\vspace{-1.2cm}
\caption{Excitation function of the nucleon chemical potential (in [GeV])
at chemical freezeout, being constrained by measured $\bar p/p$ 
ratios~\protect\cite{na4x,pbarRHIC} (filled dots).}
\vspace{-0.1cm}
\label{fig_muN}
\end{minipage}\hfill
\begin{minipage}[t]{7.5cm}
\vspace{-0.7cm}
\epsfig{file=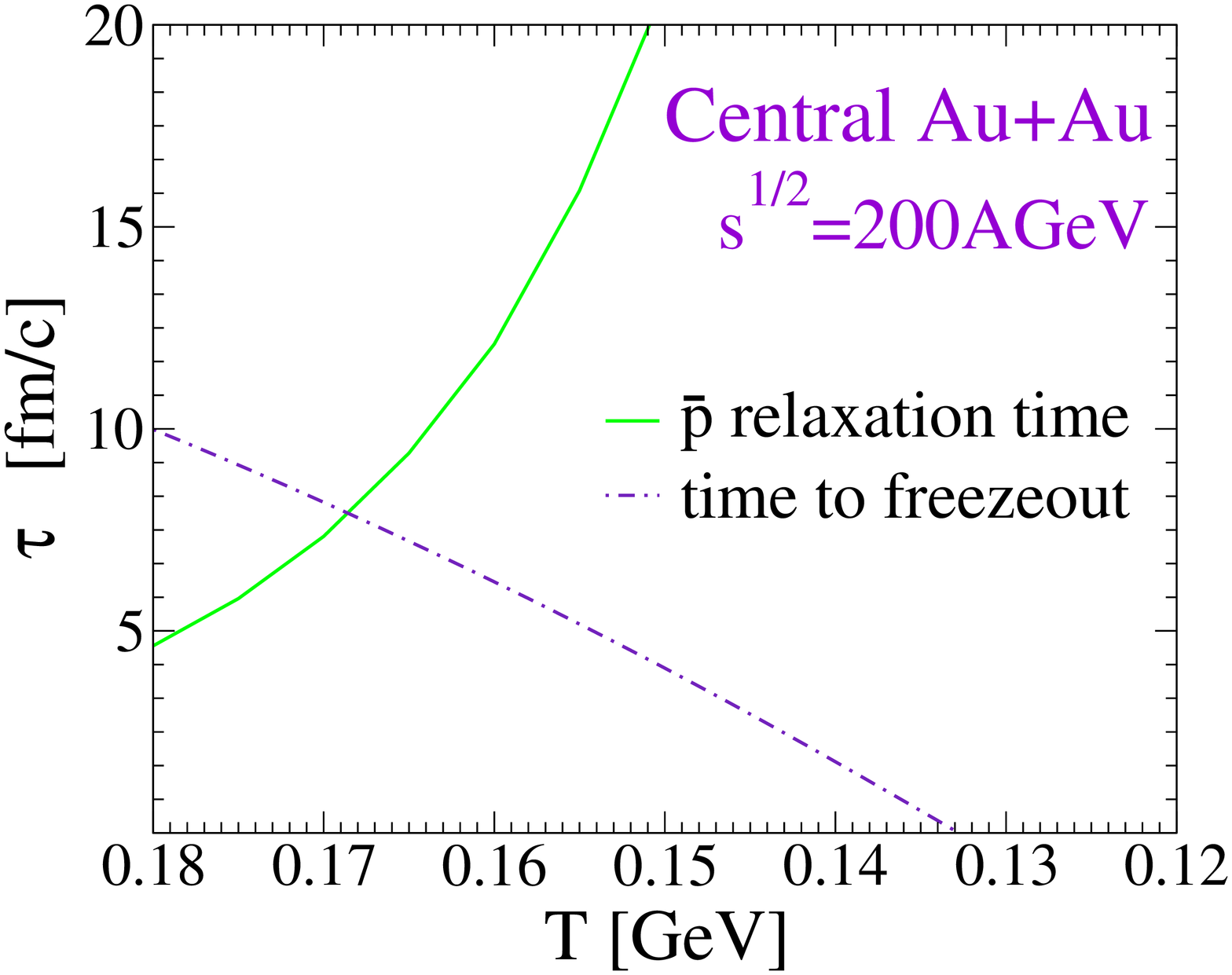,width=6.3cm,angle=-90}
\vspace{-1.2cm}
\caption{Chemical relaxation time for antiprotons (solid line)
in comparison to the remaining fireball lifetime in a thermal model
at full RHIC energy~\protect\cite{Ra01}.}
\vspace{-0.1cm}
\label{fig_tauRHIC}
\end{minipage}
\end{figure}

%%%%%%%%%%%%%%%%%%%%%%%%%%%%%%%%%%%%%%%%%%%%%%%%%%%%%%%%%%%%%%%%%%%%%%
The same approach can be applied to higher collision energies. First data
from RHIC~\cite{pbarRHIC} have shown a $\bar p/p$ ratio of $\sim$60\% at 
$\sqrt{s}$~=~130~AGeV,
implying $\mu_N^{chem}\simeq$~45~MeV (assuming $T_{chem}$~=~180~MeV), 
cf.~fig.~\ref{fig_muN}.
Extrapolating to $\sqrt{s}$~=~200~AGeV yields $\mu_N^{chem}$=27~MeV corresponding to 
$\bar p/p$=75\%. Contrary to SpS conditions, the chemical relaxation times for 
antiprotons quickly exceed the fireball lifetime (cf.~fig.~\ref{fig_tauRHIC}), 
\ie, multi-pion annihilation is not frequent enough to support the comparatively 
large antiproton abundances. Thus, the $\bar p/p$ ratio at full RHIC energy should
 be more directly associated with the chemical freezeout stage 
(the reduction in net-baryon number also leads to a much less pronounced
pion oversaturation towards thermal freezeout~\cite{Ra01}).

%%%%%%%%%%%%%%%%%%%%%%%%%%%%%%%%%%%%%%%%%%%%%%%%%%%%%%%%%%%%%%%%%%%%%%
\section{Conclusions}
%%%%%%%%%%%%%%%%%%%%%%%%%%%%%%%%%%%%%%%%%%%%%%%%%%%%%%%%%%%%%%%%%%%%%%
We have shown that the backward reaction in $p\bar p \leftrightarrow N_\pi\pi$ 
(with $N_\pi$$\simeq$6) gives important contributions to antiproton production 
in the hadronic phase of heavy-ion collisions at CERN-SpS energies. Coupled with 
realistic estimates of pion oversaturation effects, one finds a $\bar p/p$ ratio 
that with decreasing temperature deviates little from its value at chemical 
freezeout, sustaining agreement with 
experiment. This resolves the naive expectation of large annihilation losses 
towards thermal freezeout, and lends further support to an equilibrium picture 
of central Pb-Pb collisions.  Medium effects in $p\bar p \leftrightarrow N_\pi \pi$ 
reactions are neither required nor excluded,
but might be addressed within a time-dependent treatment of the rate equations.

\vspace{0.9cm}

\noindent
{\bf ACKNOWLEDGEMENT}\\
We thank K. Redlich for interesting discussion.

\vspace{0.3cm}

%%%%%%%%%%%%%%%%%%%%%%%%%%%%%%%%%%%%%%%%%%%%%%%%%%%%%%%%%%%%%%%%%%%%


\begin{thebibliography}{9}

\bibitem{pbm99}
P. Braun-Munzinger, I. Heppe and J. Stachel, Phys. Lett. {\bf B465}
(1999) 15.

\bibitem{HS98}
C.M. Hung and E.V. Shuryak, Phys. Rev. {\bf C57} (1998) 1891. 

\bibitem{WH99}
U.A. Wiedemann and U. Heinz, Phys. Rep. {\bf 319} (1999) 145.

\bibitem{RW00}
R. Rapp and J. Wambach, {\sf hep-ph/9909229} and Adv. Nucl. Phys. {\bf 25}
(2000) 1. 

\bibitem{RS00}
R. Rapp and E.V. Shuryak, {\sf hep-ph/0008326} and 
Phys. Rev. Lett. {\bf 86} (2001) 2980. 

\bibitem{GL00}
C. Greiner and S. Leupold, {\sf nucl-th/0009036}; C. Greiner, these
proceedings.   

\bibitem{H86K88}
U. Heinz, P.R. Subramanian, H. St\"ocker and W. Greiner, J. Phys. 
{\bf G12} (1986) 1237; 
P. Koch, B. M\"uller, H. St\"ocker and W. Greiner,
Mod. Phys. Lett. {\bf A3} (1998) 737.

\bibitem{na4x}
M. Kaneta for the NA44 collaboration, Nucl. Phys. {\bf A638} (1998) 419c;\\ 
G.I. Veres for the NA49 collaboration, Nucl. Phys. {\bf A661} (1999) 383c.

\bibitem{ARC}
Y. Pang, D.E. Kahana, S.H. Kahana and H. Crawford,
Phys. Rev. Lett. {\bf 78} (1997) 3418.

\bibitem{URQMD}
M. Bleicher \etal, Phys. Lett. {\bf B485} (2000) 133.

\bibitem{RW99}
R. Rapp and J. Wambach, Eur. Phys. J. {\bf A6} (1999) 415.   

\bibitem{Dov92}
C.B. Dover, T. Gutsche, M. Maruyama and A. Faessler,
Prog. Part. Nucl. Phys. {\bf 29} (1992) 87.

\bibitem{pbarRHIC}
J.W. Harris for the STAR collaboration, these proceedings;\\
W.A. Zajc for the PHENIX collaboration, these proceedings. 

\bibitem{Ra01}
R. Rapp, {\sf hep-ph/0010101} and Phys. Rev. {\bf C} (2001) in print.

\end{thebibliography}
\end{document}